\documentclass[aps,amssymb,amsmath, twocolumn, pra, superscriptaddress]{revtex4}
\usepackage{graphicx}
\usepackage{amsmath}    
\usepackage{amsfonts}
\usepackage{bm}
\usepackage{bbm}
\usepackage[colorlinks, citecolor=red]{hyperref}
\usepackage{hyperref}
\usepackage{epsfig}
\usepackage[up]{subfigure}
\usepackage{epstopdf}
\usepackage{color}
\usepackage{centernot}
\usepackage{soul}

\newcommand{\be}{\begin{equation}}
\newcommand{\ee}{\end{equation}}
\newcommand{\bc}{\begin{center}}
\newcommand{\ec}{\end{center}}
\newcommand{\bea}{\begin{eqnarray}}
\newcommand{\eea}{\end{eqnarray}}
\newcommand{\ba}{\begin{array}}
\newcommand{\ea}{\end{array}}

\newcommand{\qed}{\nobreak \ifvmode \relax \else
      \ifdim\lastskip<1.5em \hskip-\lastskip
      \hskip1.5em plus0em minus0.5em \fi \nobreak
      \vrule height0.75em width0.5em depth0.25em\fi}

\begin{document}

\title{Single photons in an imperfect array of beam-splitters: Interplay between percolation, backscattering and transient localization}
\author{C.~M.~Chandrashekar}
\email{c.madaiah@oist.jp}
\affiliation{Quantum Systems Unit, Okinawa Institute of Science and Technology Graduate University, Okinawa, Japan}
\author{S.~Melville}
\affiliation{The Queen's College, University of Oxford, United Kingdom}
\author{Th.~Busch}
\affiliation{Quantum Systems Unit, Okinawa Institute of Science and Technology Graduate University, Okinawa, Japan}


\begin{abstract}
\noindent
Photons in optical networks can be used in multi-path interferometry and various quantum information processing and communication protocols. Large networks, however, are often not free from defects, which can appear randomly between the lattice sites and are caused either by production faults or deliberate introduction. In this work we present numerical simulations of the behaviour of a single photon injected into a regular lattice of beam-splitting components in the presence of defects that cause perfect backward reflections. We find that the photon dynamics is quickly dominated by the backscattering processes, and a small fraction of reflectors in the paths of the beam-splitting array strongly affects the percolation probability of the photon. We carefully examine such systems and show an  interesting interplay between the probabilities of percolation, backscattering and temporary localization. We also discuss the sensitivity of these probabilities to lattice size, timescale, injection point, fraction of reflectors and boundary conditions.
\end{abstract}

\maketitle
\section{Introduction}

Recent developments in experimental techniques have allowed the realisation and study of many complex photonic systems such as multipath, multiphoton interferometers that exhibit high fidelity quantum interference\,\cite{PCR08, M+09, PLM10, L+10, PLP11, S12}. This stems from, and also stimulates, a great deal of interest in using photons as information carriers for various quantum information processing and communication protocols\,\cite{YSS05, PMO09, SCP10, RBH11, B+13, S+13, TDH13}.  However, building the large optical networks for photon propagation required by some of these protocols is not an easy task and imperfections in the coupling between different sections of a network can appear. It is therefore important to discuss and simulate simple toy models of single photon propagation in an irregular array of beam-splitters, in order to achieve a better understanding of how proposed large optical networks might behave in practice.
\par
Here we present a numerical study of the behaviour of a single photon injected into a regular lattice of beam splitting components (modelling the network), in which we allow for perfect reflections to occur between a certain fraction of the lattice sites (modelling the system defects, or an intentional feature of the network). Though the the presence of the reflectors  introduces irregular paths for photon propagation, the operation at each lattice site is considered to be an ideal lossless beam-splitter, where the input and output operators are related by a unitary transformation. We find that the photon is confined within a lattice of size $N\times N$ over timescales proportional to $N$, but that these vary considerably with factors such as the injection point and the boundary conditions of the lattice, which we choose as either reflective or absorptive. This allows for temporary localization of the photon within the lattice network and, as time progresses, there is a non-trivial tradeoff between the probabilities for localization, percolation, and backscattering.  
\par
Our presentation is organised as follows. In section\,\ref{Photonarray} we define the dynamics of a photon in a completely connected array of beam-splitters and in section\,\ref{Photonarrayref} we simulate the dynamics in the presence of a number of reflectors between adjacent beam-splitters. We then calculate the probabilities of percolation, backscattering and temporary localization and conclude with a discussion of the results in section\,\ref{conclusion}.

\section{Photon propagation in a regular array of beam-splitters}
\label{Photonarray}

\begin{figure}[tb]
\includegraphics[width=0.5\textwidth]{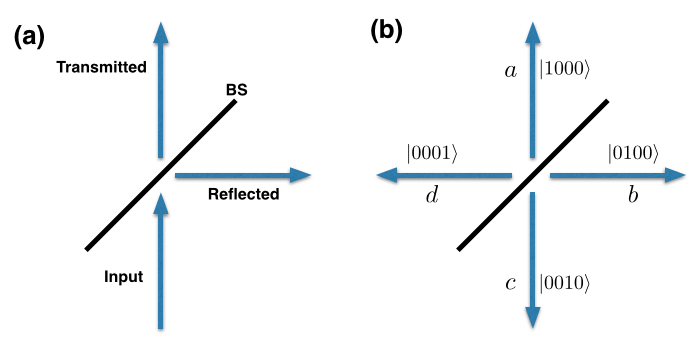}
\caption{(a) Schematic of a beam-splitter (BS) with the output paths (reflected and transmitted) indicated for one of the possible input states. (b) All possible photon modes outgoing from the beam-splitter.  In the percolation direction the input state $|1000\rangle$ leads to the output states $|1000\rangle$ and $|0100\rangle$ and the input state $|0100\rangle$ leads to $|0100\rangle$ and $|1000\rangle$. In the backscattering direction the input state $|0010\rangle$ leads to the output states $|0010\rangle$ and $|0001\rangle$ and the input state $|0001\rangle$ leads to $|0001\rangle$ and $|0010\rangle$\label{fig:1}}
\end{figure}

A photon incident on a beam-splitter can be written as the Fock state $| n_{a}, n_{b}, n_{c}, n_{d} \rangle$. For a single photon, $n_{a} + n_{b} + n_{c} + n_{d} = 1$ with each $n$ being an integer and the indices $a, b, c, d$ specifying the four beam-splitter arms. In figure\,\ref{fig:1}(a) we show a schematic of a photon impinging on a beam-splitter and indicate the corresponding transmitting and reflecting paths. In figure\,\ref{fig:1}(b) we define the four arms of the beam-splitter as $a$, $b$, $c$, and $d$ and indicate the corresponding Fock states for a photon travelling in one of the associated modes.
This allows to define annihilation operators $\hat a, \hat b, \hat c, \hat d$, such that
\begin{align}
  &\hat a | 1 0 0 0 \rangle  = | 0 0 0 0 \rangle, \quad 
   \hat a^\dagger | 0 0 0 0 \rangle = | 1 0 0 0 \rangle  \\
  &[ \hat a, \hat a^\dagger ] = 1; \quad
  [\hat a, \hat b] = [\hat a, \hat c]  = [\hat a, \hat d] = 0,
\end{align}
and analogously for the other three operators ($\hat b,\hat c,\hat d$) corresponding to the remaining three indices ($n_{b},n_{c},n_{d}$). Thus the action of a beam-splitter on a photon may be regarded as the action of the effective Hamiltonian
\begin{equation}
\label{eq:2}
\begin{split}
H &= \frac{1}{\sqrt{2}} \left ( \hat a^\dagger - i \hat b^\dagger \right) \hat a  +  \frac{1}{\sqrt{2}} \left(  \hat b^\dagger -i  \hat a^\dagger \right) \hat b \\
   & \;\;\; +  \frac{1}{\sqrt{2}} \left(\hat  c^\dagger - i \hat d^\dagger \right) \hat  c  + \frac{1}{\sqrt{2}}  \left( \hat  d^\dagger - i \hat  c^\dagger   \right) \hat  d\;,
\end{split}
\end{equation}
where the factor of $i$ accounts for a phase shift of $\pi$ during reflection.  
\par

\begin{figure}[tb]
\includegraphics[width=0.45\textwidth]{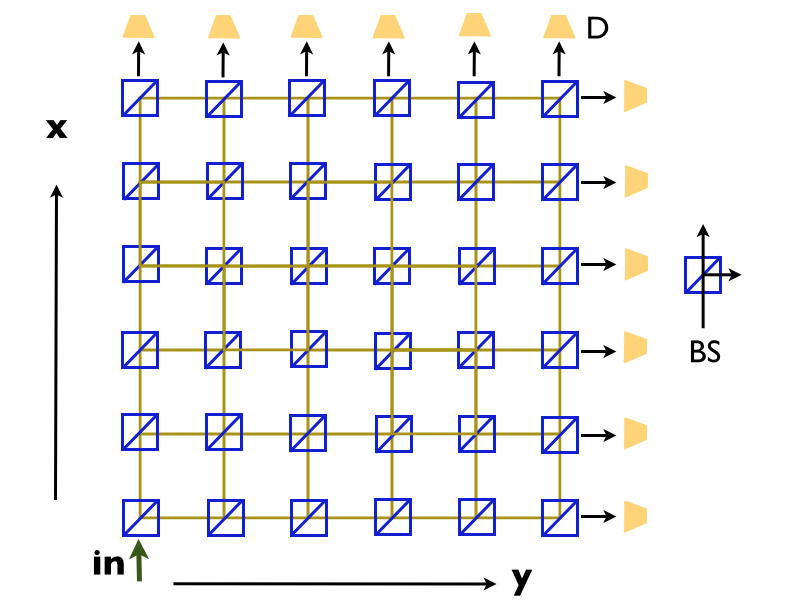}
\caption{Schematic of an array of beam-splitters arranged in a square lattice with detectors (D) at all possible output ports, which register the photon once it has moved through  the array.  The small graph at the right hand side indicates the possible paths for a photon entering in $|1000\rangle$. \label{fig:2}}
\end{figure}

We will now consider an array of  beam-splitters, each positioned at the vertices of a square lattice and labelled by  $(x,y)$ (see figure\,\ref{fig:2}). Initially, a single photon is injected at  $(x, y) = (1, 1)$ in state $|1000\rangle$ and we can describe its dynamics using the product basis $ | n_{a}, n_{b}, n_{c}, n_{d} \rangle   \otimes \mathcal{H}_{x,y}$, where $\mathcal{H}_{x,y}$ is the position Hilbert space. Therefore, the initial state at the injection point as shown in figure\,\ref{fig:2} will be given by
\be
\label{eqn:InitialCondition}
|\Psi(t=0) \rangle = |1000\rangle \otimes |x =1, y=1\rangle.
\ee
The action of the beam-splitting operator, which acts only on the Fock state $|n_a, n_b, n_c, n_d\rangle$  and leaves the position states unchanged, will be $H$ (equation\,(\ref{eq:2})), 
and the evolution of the position state is given by the shift operation
\bea
\begin{split}
\label{eqn:S1}
S &= \sum_{(x,y)} |1000 \rangle \langle1000| \otimes |x+1,y  \rangle \langle x,y| \\
& ~~~~~~+ |0100  \rangle \langle 0100|   \otimes |x,y+1  \rangle \langle x,y| \\
& ~~~~~~+ |0010  \rangle \langle 0010|   \otimes |x-1,y  \rangle \langle x,y| \\
& ~~~~~~+ |0001  \rangle \langle 0001|   \otimes |x,y-1  \rangle \langle x,y|.
\end{split}
\eea
Hence the successive action of $H$ and $S$ on the product state $| n_{a}, n_{b}, n_{c}, n_{d} \rangle   \otimes  |x,y \rangle $ advances the system one time step, and after $t$ steps the state of the photon is given by
\begin{equation}
\label{eqn:TimeEvolution}
| \Psi ( t ) \rangle =  \left[ S \left( H \otimes {\mathbbm 1} \right) \right] ^{t} | \Psi ( t=0 ) \rangle.
\end{equation}
In this regular evolution the photon will never be scattered into the modes $|0010\rangle$ and $|0001\rangle$ therefore it can only exit at the upper and right-hand side edges of the lattice. We call this forward propagation. If we define the time required for the photon to travel between two beam-splitters as unity, the total probability for the photon to reach an edge of a lattice of size $N\times N$  is $P(t) =0$ for $t \leq N$,  $P(t) =1$ for $t > 2N$ and $0 \leq P(t) \leq 1$ for any time $ N < t < 2N$.  

\section{Photon propagation in an array of beam-splitters with backward reflectors}
\label{Photonarrayref}

Backward reflection and loss of photons between the lattice site are two of the most fundamental processes that can affect the forward propagation of a photon in an array of beam-splitting components.  In this section we will discuss the additional effects that appear when a certain number of backward reflectors are introduced into the path. While the results are specific to the setup, the treatment we present can serve as a general framework for other forms of irregularities in the path of the photons.

\begin{figure}[tb]
\includegraphics[width=0.5\textwidth]{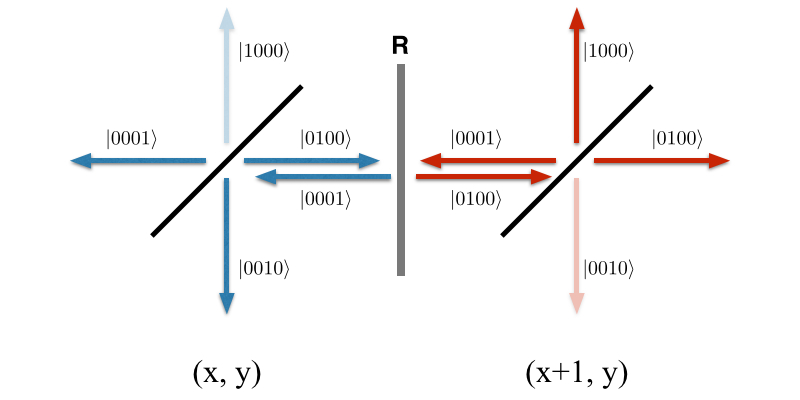}
\caption{ Schematic of two neighbouring beam-splitters with a reflector (R) in the connecting path. The initial output state from the blue beam-splitter (left-hand side) is  $|0100\rangle$ and the one from the red beam-splitter (right-hand side) is $|0001\rangle$. Thus $k_a (x,y)=1$, $k_c (x+1,y)=1$. \label{fig:3}}
\end{figure}

\begin{figure}[tb]
\includegraphics[width=0.45\textwidth]{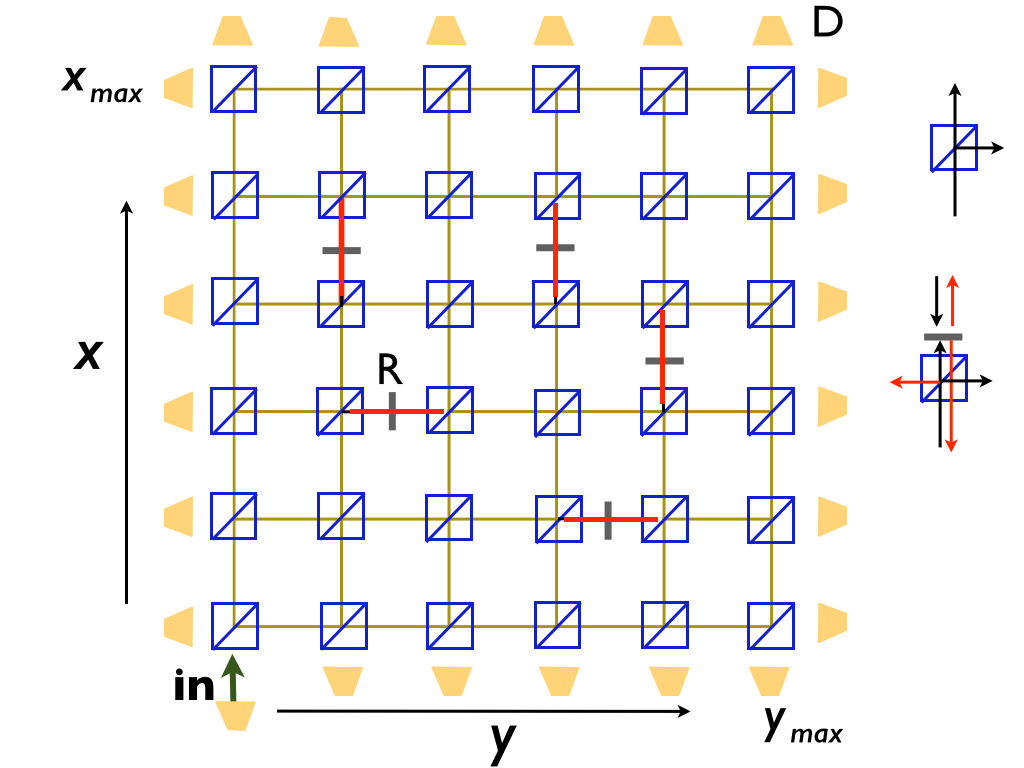}
\caption{ Schematic of the array of beam-splitters in a square lattice with impurities given by perfect reflectors. Photon detectors along $(x=1,y)$ and $(x,y=1)$ will register the backscattering of the photon due to the presence of the reflectors. 
The small graphs at the right hand side indicate the possible paths for a photon at each vertex.
\label{fig:4}}
\end{figure}
In figure\,\ref{fig:3} we show the effect a reflector, positioned between two beam-splitters, has on the path of a photon and in figure\,\ref{fig:4} a schematic of an array of beam-splitters interspersed with a number of reflectors is given. In order to model the effect of perfect reflection at the beam-splitters, we consider the 
initial state at the injection point to be given by equation\,\eqref{eqn:InitialCondition}. Note that for symmetry reasons the results obtained below also hold for a photon initially entering in mode $|0100\rangle$. For all \emph{completely connected} vertices the Hamiltonian $H$ is given in equation\,(\ref{eq:2}) and can be written as,
\begin{equation}
H = \frac{1}{\sqrt{2}}
\begin{pmatrix}
\hat a & \hat b & \hat c & \hat d\, 
\end{pmatrix}
\begin{pmatrix}
1 & -i & 0 & 0 \\
-i & 1 & 0 & 0 \\
0 & 0 & 1 & -i \\
0 & 0 & -i & 1 \\
\end{pmatrix}
\begin{pmatrix}
\hat a^\dagger \\ \hat b^\dagger \\ \hat c^\dagger \\ \hat d^\dagger \\
\end{pmatrix}\;.
\end{equation}
When a reflector is present in an arm between two vertices the general Hamiltonian can be written in the form
\begin{equation}
H =  \frac{1}{\sqrt{2}}
\begin{pmatrix}
\hat a & \hat b & \hat c & \hat d\,
\end{pmatrix}
R
\begin{pmatrix}
\hat a^\dagger  \\ \hat b^\dagger  \\ \hat c^\dagger  \\ \hat d^\dagger  
\end{pmatrix}\;,
\end{equation}
where $R$ is given by
\begin{equation}
R = \begin{pmatrix}
~~~1-k_a & -i(1-k_b) & -ik_a &~~ k_b \\
-i(1-k_a) & ~~~~1-k_b &~~~~ k_a & -ik_b \\
-ik_c & ~~~~k_d & ~~~~1-k_c & -i(1-k_d) \\
~~k_c &-ik_d & -i(1-k_c) & ~~~1-k_d \\
\end{pmatrix}
\end{equation}
with $k_n = 0$ if the $n^\mathsf{th}$ arm is open and $k_n = 1$ if the $n^\mathsf{th}$ arm contains a reflector. 
The corresponding shift operator is then
\begin{align}
\label{eqn:S}
S = \sum_{(x,y)}&\;  |1000 \rangle \langle1000| \otimes |x+(1-k_a),y  \rangle \langle x,y|  \nonumber\\ 
 +&\; |0100  \rangle \langle 0100| \otimes |x,y+(1-k_b)  \rangle \langle x,y|   
 \nonumber\\ 
 +& \;|0010   \rangle \langle 0010| \otimes |x-(1-k_c),y  \rangle \langle x,y|  
 \nonumber \\
 +& \;|0001  \rangle \langle 0001| \otimes | x,y-(1-k_d)  \rangle \langle x,y| ,
\end{align}
and the system evolves according to the modified equivalent of equation\,(\ref{eqn:TimeEvolution}). This  ensures that a photon in, for example, the $|0100\rangle$ mode will scatter into the $|0001\rangle$ mode and acquire a phase shift of $\pi$ when hitting a reflector. This photon will also be unaffected by $S$, so that it encounters the same beam-splitter a second time at the subsequent time step. The distribution of reflectors in the lattice is given by a consistent set of $k_i (x,y)$ such that $k_a (x,y) = k_c (x+1,y) $, etc.
\begin{figure}[tb]
\begin{center}
  \includegraphics[width=0.45\textwidth]{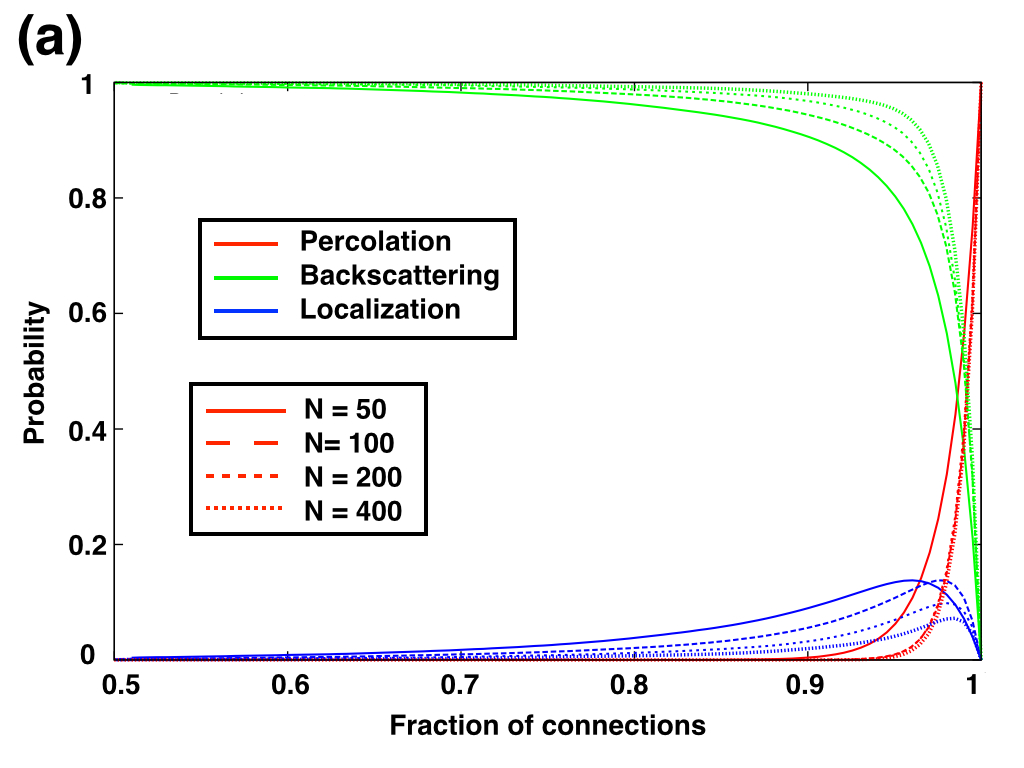}
  \includegraphics[width=0.45\textwidth]{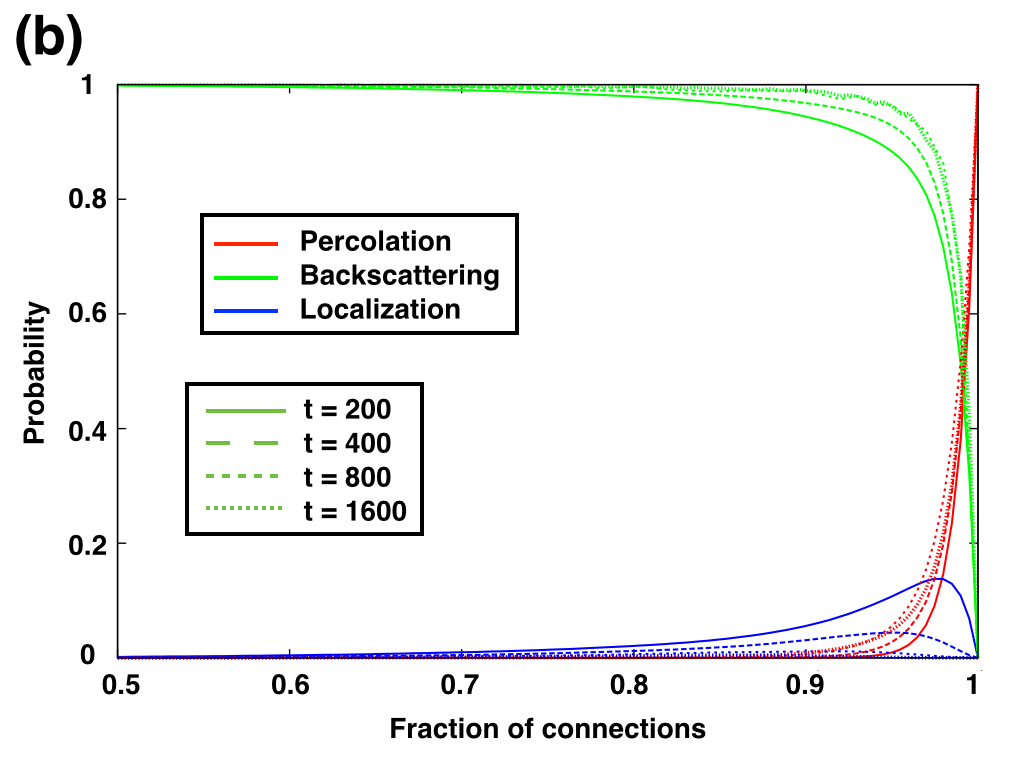}
\end{center}
\caption{Probability of photon percolation, backscattering and temporary localization as a function of the fraction of connections between adjacent beam-splitters. (a) Probabilities for lattices of different sizes, $N \times N$, where $N=50, 100, 200$ and $400$ are shown at time $t=2N$.   (b) Probabilities for a lattice of size $N=100$ for different times. Strong backscattering is clearly visible until the fraction of connections between the adjacent beam-splitters is close to unity.\label{fig:5}}
\end{figure}
\par
During this evolution the reflections can lead to backscattering of the photon (i.e.~scattering into the modes $|0010\rangle$ and $|0001\rangle$), which opens the possibility for the photon to exit along the lattice edges on the left and the bottom. Additionally, sufficiently nearby groups of such reflectors can lead to temporary localization of the photon in the lattice.  Therefore, in addition to the percolation probability, the system is characterised by  probabilities for backscattering and localization.  Assuming an arrangement of detectors as shown in figure\,\ref{fig:4}, {\it percolation} corresponds to the photon exiting the lattice from either of the edges $(x_\text{max}, ~y)$ or $(x, ~y_\text{max})$, {\it backscattering} corresponds to exiting the lattice from the edges along $(1, ~y)$ and $(x, ~1)$, and {\it localization} corresponds to temporary confinement within the lattice for times $t\ge2N$. Since all possible photon paths are reversible, localization is of course only transient. 
For the initial state given in equation\,\eqref{eqn:InitialCondition}, i.e.~injecting a single photon at one of the corners of the lattice, we show in figure\,\ref{fig:5} the probabilities of percolation, backscattering and temporary localization as a function of the fraction of connections between adjacent beam-splitters that are not disturbed by a reflector. These probabilities are obtained after averaging over a large number of realizations.
\par
The probabilities for lattices of different sizes $N \times N$, where $N=50, 100, 200$ and $400$, at time $t=2 N$ are shown in figure\,\ref{fig:5}(a). One can note that the probability for backscattering  dominates until the fraction of connections between the adjacent beam-splitters is close to unity and one can think of the fraction at which a finite probability for percolation appears as the analogue to the classical percolation threshold\,\cite{K73+, SA94, CB13}. This behaviour can be easily understood by realising that encountering a reflector once leads to scattering into the modes that lead to backscattering, and encountering a second reflector is necessary to scatter into the percolation modes again. Since the injection point is located at the corner of the network furthest away from any detectors for percolation, reflection early on during the propagation process lead to the  domination of the backscattering probability.
When the fraction of connections is closer to unity, but before the steep increase in percolation probability dominates, temporary confinement of the photon within the lattice can be seen. This indicates that, while a large number of reflectors leads to quick expulsion of the photon along the sides with $(1, y)$ and $(x, 1)$, a decreasing number allows for geometries in which the photon bounces around inside the lattice for a long time. A large fraction of good connections between the beam-splitting components is therefore required for the photon to percolate across an array of beam-splitters. From figure\,\ref{fig:5}(a), one can also note that the lattice size (the number of beam-splitters) has only a weak influence on these probabilities. The probabilities for a lattice with $N=100$ for different times are shown in figure\,\ref{fig:5}(b) and one can see the probability of temporary localization decreasing with time, as expected. 
\begin{figure}[tb]
\includegraphics[width=0.45\textwidth]{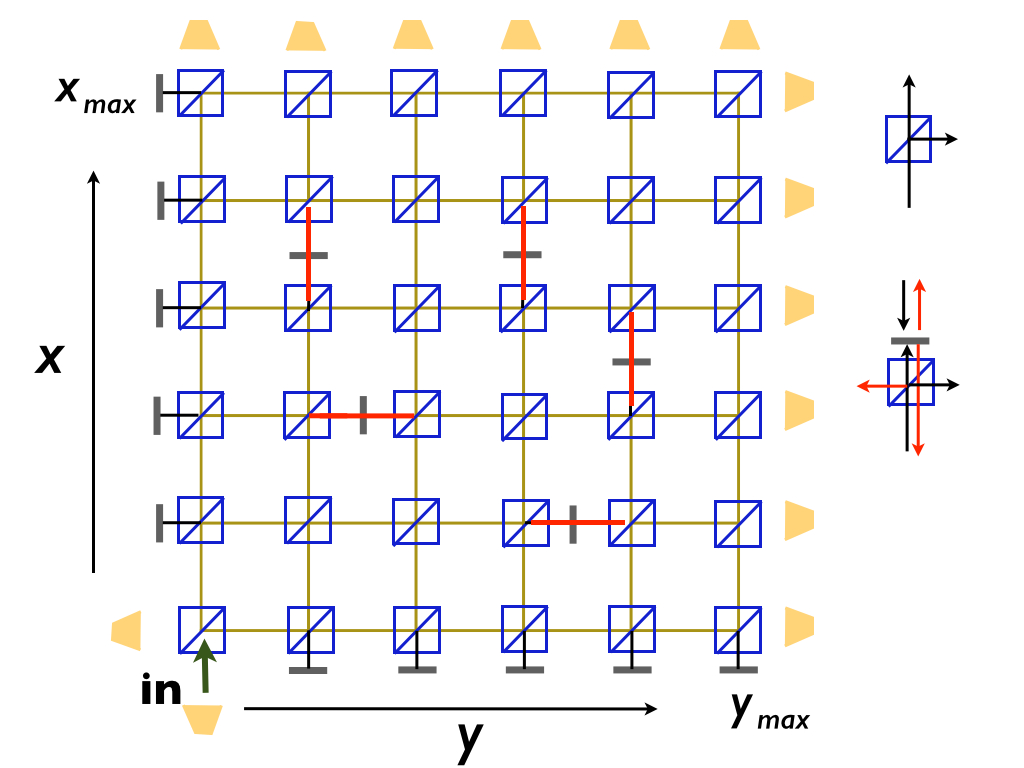}
  \caption{Schematic of an array of beam-splitters in a square lattice interspersed with a small number of perfectly reflecting surfaces and reflecting boundaries. A photon backscattered along the injection side of the lattice is fed back to the lattice due to reflectors placed along these sides, except at the injection point. \label{fig:6}}
\end{figure}
\begin{figure}[tb]
\begin{center}
  \includegraphics[width=0.45\textwidth]{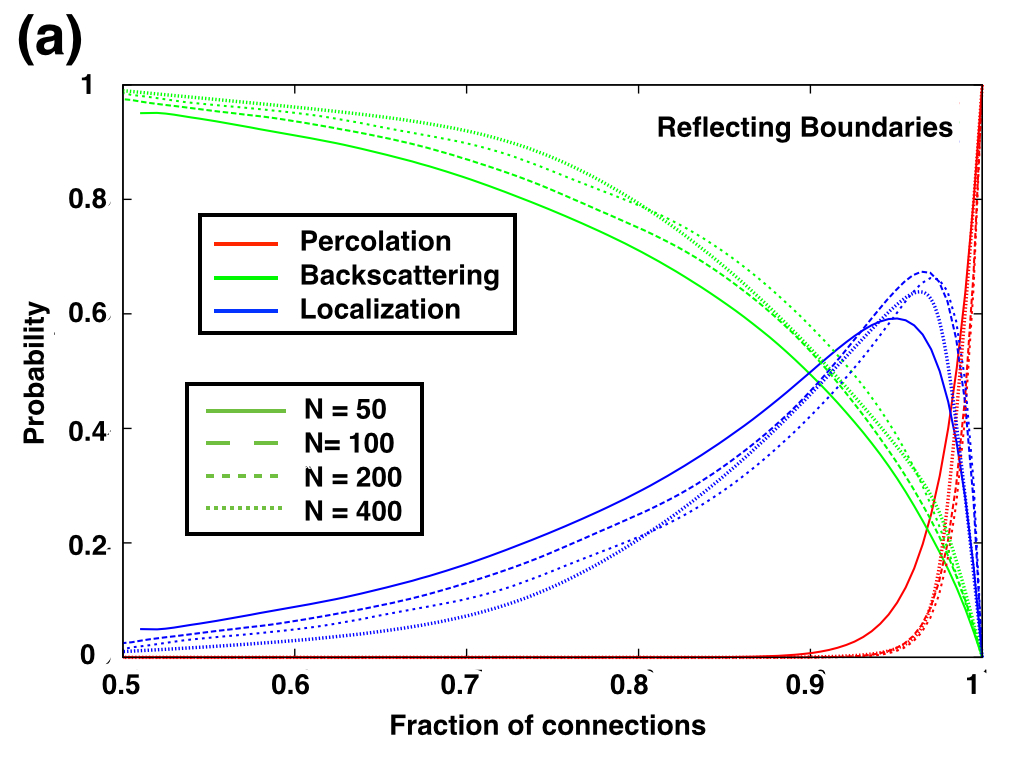}
 \includegraphics[width=0.45\textwidth]{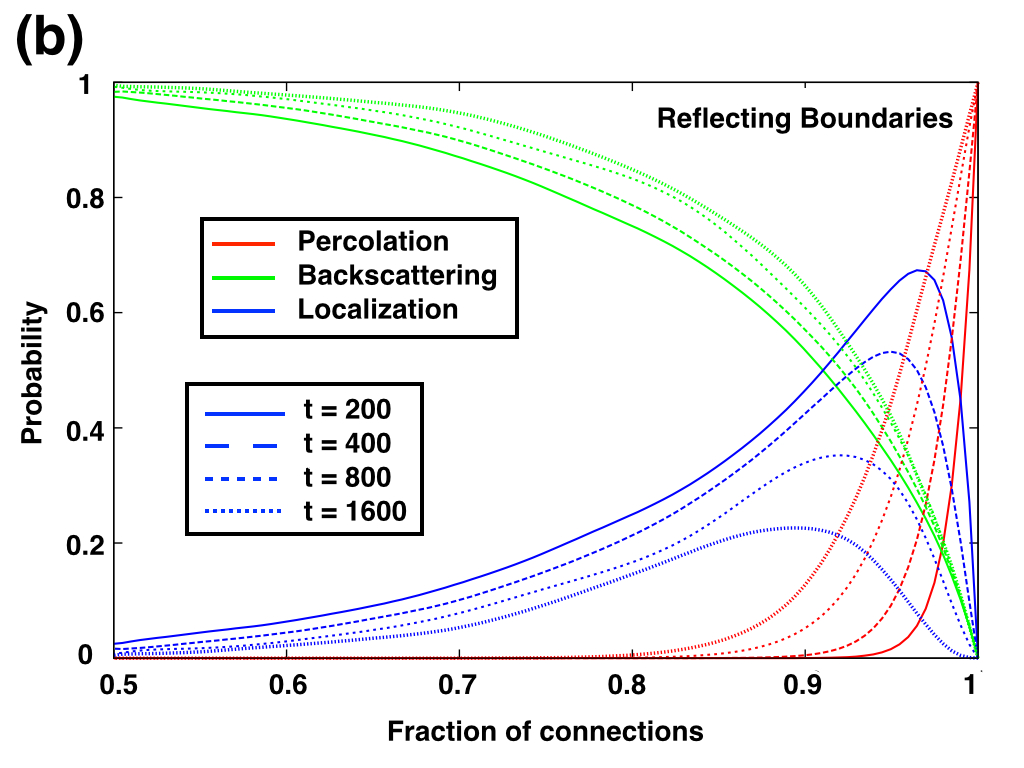}
 \end{center}
\caption{Probability of photon percolation, backscattering and temporary localization as a function of fraction of connections between the adjacent beam-splitters for the situation where a backscattered photon is fed back to the lattice at the edges. (a) Probabilities for lattices of different sizes, $N \times N$, where $N=50, 100, 200$ and $400$ are shown at time $t=2N$.  (b) Probabilities for a lattice with $N=100$ for different times. \label{fig:7}}
\end{figure}
\vskip 0.2in
\par
\begin{figure}[tb]
\includegraphics[width=0.45\textwidth]{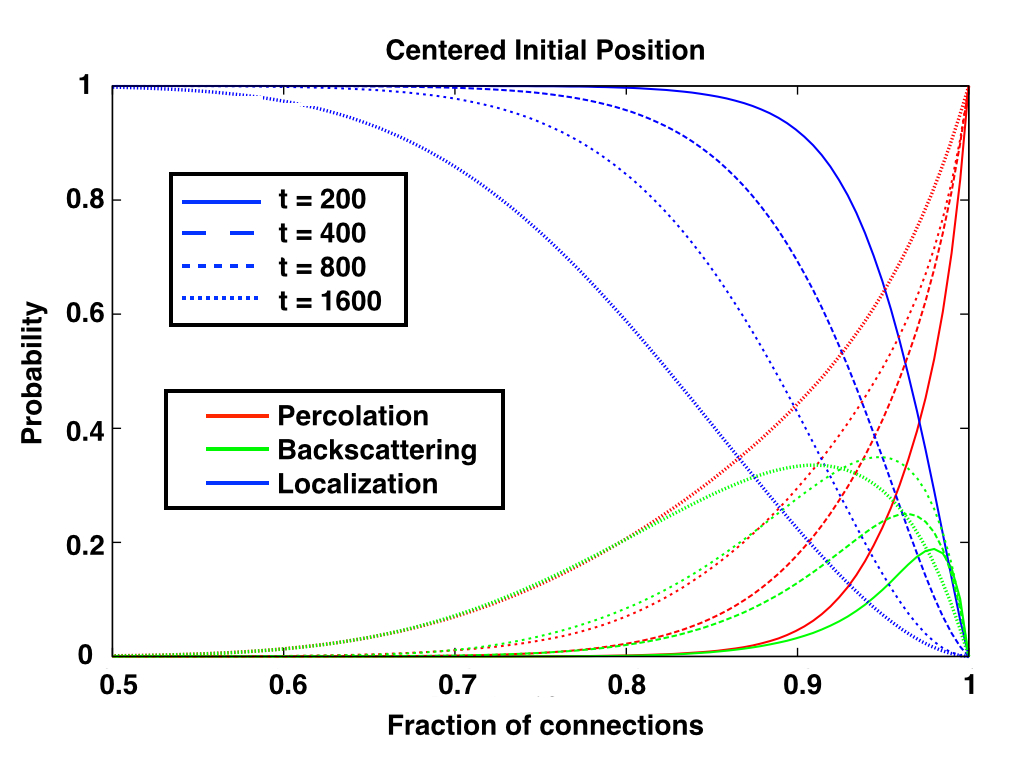}
\caption{Probability of photon percolation, backscattering and temporary localization as a function of fraction of connections between the adjacent beam-splitters in the absence of reflecting boundaries and with the photon incident at the center of the lattice array. The probabilities for a lattice with $N=100$ for different times are shown.\label{fig:8} }
\end{figure}
\par
The interplay between backscattering, localization and percolation can be changed by introducing reflecting edges in the backscattering direction and allowing backscattered photons to only exit at the injection point ($x=1,y=1$) (see figure\,\ref{fig:6}). Unsurprisingly one can see from figure\,\ref{fig:7}(a), where we show the probabilities for different lattices sizes, that at $t=2 N$ backscattering is reduced and instead an increase in temporary localization is observed compared to the situation when reflecting edges are absent (see figure\,\ref{fig:5}(a)). Backscattering is still significant though, since photons scattered early on in the percolation process have a high probability to exit through the entry beam-splitter and this probability is further increased by the coherent backscattering\,\cite{Akkermans:86}. In figure\,\ref{fig:7}(b) we show the probabilities for different times, and find that the probability for localization monotonically decreases while both, the backscattering and percolation probability rise. One can again note that the dependence on the lattice size (the number of beam-splitters) has only a weak influence on the probabilities. 
\par
Comparing both cases above one can  note that for the identical initial condition given by equation\,(\ref{eqn:InitialCondition}), the asymptotic behaviour is identical: as the fraction of connections goes to unity the percolation probability goes to one, whereas for a fraction of connections around 0.5, backscattering has a probability of one. While different initial conditions exhibit qualitatively the same interplay between transient localization and percolation (with strong dependence on detector placement and weak dependence on lattice size), the general asymptotic behaviour will change. An example of this is shown in figure\,\ref{fig:8} for the situation without reflecting boundary conditions and where we have chosen $|\psi(t=0)\rangle = |1000\rangle \otimes |\frac{1}{2}x_\text{max}, \frac{1}{2}y_\text{max} \rangle$.
For consistency we will again define percolation as exiting the lattice in the modes $|1000\rangle$ or $|0100\rangle$ and backscattering as having encountered an odd number of reflectors before leaving the lattice in the modes $|0010\rangle$ or $|0001\rangle$. In figure\,\ref{fig:8} we show the resulting probabilities and one can note that the percolation probability is almost same as the backscattering probability 
until the fraction of connection gets closer to unity ($0.8$ for $t =1600$ when $N=100$). After that the backscattering probability decreases to zero and the percolation probability rises to one, as the photon can transverse the upper right quarter of the network most of the times without encountering a reflector. Fraction of connections smaller than $0.5$  results in \emph{localization} with probability one.  Unlike the transient localization for the model with the injection point at one of the corners of the lattice, the localization for injection at the middle of the lattice is a permanent localization, due to the absence of a detector close to the injection point.

\section{Discussion and conclusion}
\label{conclusion}
In this work we have modelled a large optical network consisting of a regular array of beam-splitters, and considered the effects stemming from randomly introduced reflective defects. The presence of these defects has a significant influence on the transport properties of the system - with the percolation probability for a photon decaying rapidly even for only a small percentage of defective paths ($\sim 10\%$). We have also found the existence of a transient `localised' state, which confines the photon within the lattice over finite timescales.
\par 
In region of small percentages of defects, an interesting interplay between the three possible scenarios takes place: the photon percolates forward, the photon backscatters, or the photon remains within the lattice. These relative probabilities are fairly insensitive to changes in the lattice size, but vary significantly if the distribution of detectors around the lattice is altered (by replacing some detectors with reflectors, feeding those photons back into the lattice). With fewer detectors around the lattice edges, the localization probability is finite over a much longer timescales, before giving way to both, backscattering and percolation. If the injection point is near a particular lattice edge, a large probability for the photon to exit the lattice via this edge exists (backscattering processes dominate), and if the injection point is far from a lattice edge, long-lived localization can be seen.
\par 
The implication for large optical networks is that even small fractions of reflective defects will significantly alter the path taken by the photon through the system. Therefore, quantum communication systems using optical networks  will be very sensitive to defects and require additional strategies to combat imperfections. These could, for example, consist of the suitable use of additional reflectors to feed stray photon amplitudes back into the system. The study of multipath interferometer or large optical networks are therefore very valuable to identify the percentages of defective components a system can tolerate and  to test ideas to correct them in oder to obtain reliable devices.

\acknowledgments 
SM would like to thank OIST Graduate University for the support for summer internship during which this work was carried out.


\end{document}